\documentclass[11pt]{article}
\usepackage{blois2002,epsfig}
\def\ra{\rightarrow}

\def\be{\begin{equation}}
\def\ee{\end{equation}}
\def\bea{\begin{eqnarray}}
\def\eea{\end{eqnarray}}

\newcommand{\lbl}[1]{\label{eq:#1}}
\newcommand{ \rf}[1]{(\ref{eq:#1})}

\newcommand{\noi}{\noindent}

\newcommand{\Ra}{\Rightarrow}

\newcommand{\lesssim}{ {\
\lower-1.2pt\vbox{\hbox{\rlap{$<$}\lower5pt\vbox{\hbox{$\sim$}}}}\ } 
}
\newcommand{\gtrsim}{ {\
\lower-1.2pt\vbox{\hbox{\rlap{$>$}\lower5pt\vbox{\hbox{$\sim$}}}}\ } 
} 

\newcommand{\cA}{{\cal A}}

\newcommand{\cC}{{\cal C}}

\newcommand{\cH}{{\cal H}}

\newcommand{\cL}{{\cal L}}

\newcommand{\cO}{{\cal O}}
\newcommand{\cP}{{\cal P}}
\newcommand{\cR}{{\cal R}}

\newcommand{\cW}{{\cal W}}

\newcommand{\Imm}{\mbox{\rm Im}}
\newcommand{\Ree}{\mbox{\rm Re}}

\newcommand{\tr}{\mbox{\rm tr}}

\newcommand{\MeV}{\mbox{\rm MeV}}
\newcommand{\GeV}{\mbox{\rm GeV}}

\newcommand{\annd}{\mbox{\rm and}}

\newcommand{\GF}{G_{\mbox{\rm {\small F}}}}

\newcommand{\gL}{\frac{1-\gamma_{5}}{2}}

\newcommand{\eff}{\mbox{\rm eff}}

\newcommand{\QCD}{\mbox{\rm {\footnotesize QCD}}}

\newcommand{\stern}{\langle\bar{\psi}\psi\rangle}

\input epsf

\begin{document}
\vspace*{4cm}
\title{THEORETICAL ASPECTS OF RARE KAON DECAYS~\footnote{Talk given by E. de
Rafael}}

\author{David Greynat and Eduardo de Rafael }

\address{CPT, CNRS--Luminy, Marseille, France}

\maketitle\abstracts{ Most of the analytic approaches which are used at present to
understand Kaon decays, get
their inspiration from QCD in the limit of a large number of colours $N_c$.
After a general overview, we illustrate this with a recent application to
the evaluation of the process $K_L\ra\mu^+ \mu^-$ in the Standard Model.}

\section{INTRODUCTION}
In the Standard Model, the electroweak interactions of hadrons at
very low energies are conveniently described by an effective chiral
Lagrangian, which has as active degrees of freedom the low--lying
$SU(3)$ octet of pseudoscalar particles, plus leptons and photons.
The underlying theory is a  $SU(3)_{C}\times
SU(2)_{L}\times U(1)_{W}$ gauge theory which is formulated in terms
of quarks, gluons and leptons, together with the massive gauge
fields of  the electroweak interactions and the hitherto unobserved
Higgs particle. Going from the underlying Lagrangian to the 
effective chiral Lagrangian is a typical renormalization group
problem.  It has been possible to integrate the heavy degrees of
freedom of the underlying theory, in the presence of the strong
interactions, perturbatively, thanks to the asymptotic freedom
property of the $SU(3)$--QCD sector of the theory. This brings us down
to an effective  field theory which consists of the QCD Lagrangian
with the $u$, $d$,
$s$ quarks still active, plus a string of four quark operators and
mixed quark--lepton
operators, modulated by coefficients which are functions of the
masses of the fields which have been integrated out and the scale
$\mu$ of whatever renormalization scheme has been used to
carry out this integration. We are still left 
with the evolution from this effective field theory, appropriate at
intermediate scales of the order of a few $\GeV$, down to an effective
Lagrangian description in terms of the  low--lying pseudoscalar particles
which are the Goldstone modes associated with the spontaneous symmetry
breaking of chiral--$SU(3)$ in the light quark sector. The dynamical
description of this evolution involves genuinely non--perturbative
phenomena. There has been recent progress in approaching this last step,
using analytic methods  formulated
within the context of QCD in the limit of a large number of
colours $N_c$ ({$\QCD_{\infty}$), (\it see e.g. ref~\cite{EdeR02} for a
recent review.})

The strong and electroweak interactions of the Goldstone modes at
very low energies are described by an effective Lagrangian which has
terms with an increasing number of derivatives (and quark masses if
explicit chiral symmetry breaking is taken into account.) Typical terms
of the chiral Lagrangian are

\bea\lbl{chiral}
 \cL_{\eff}  & =  & \underbrace{\frac{1}4F_{0}^2\
\tr\left(D_{\mu}U
D^{\mu}U^{\dagger}\right)}_{ 
\pi\pi\ra\pi\pi\,,\quad K\ra\pi e\nu
}  +\underbrace{L_{10}
\tr\left(
U^{\dagger}F_{R\mu\nu}
UF_{L}^{\mu\nu}\right)}_{ \pi\ra
e\nu\gamma }+\cdots \\ 
&  &  
+\underbrace{e^{2}C
\tr\left(
Q_{R}UQ_{L}U^{\dagger}\right)}_{
-e^2 C\frac{2}{F_{0}^{2}}\left(\pi^{+}\pi^{-}+K^{+}K^{-}
\right)}+\cdots \lbl{c} \\ & &
-\underbrace{\frac{\GF}{\sqrt{2}}V_{ud}V^{*}_{us} 
\ g_{\underline{8}}F_{0}^{4}\left(D_{\mu}U
D^{\mu}U^{\dagger}\right)_{23}}_{
{K\ra\pi\pi\,, \quad K\ra\pi\pi\pi}}+\cdots\,, \lbl{g8}
\eea

\noi
where $U$ is a $3\times 3$ unitary matrix in flavour space which
collects the Goldstone fields and which under chiral rotations
transforms as $U\ra V_{R}U V_{L}^{\dagger}$; $D_{\mu}U$ denotes the
covariant derivative in the presence of external
vector and axial--vector sources. The first term in the first line is the
lowest order term in the sector of the strong interactions~\cite{We79},
$F_0$ is the pion--decay coupling constant in the chiral limit where the
light quark masses $u$, $d$, $s$ are neglected ($F_0\simeq 90~\MeV$);
the second term shows one of the couplings at
$\cO(p^4)$~\cite{GL84,GL85}; the second line shows the lowest order
term which appears when  photons and $Z's$ are integrated out ($Q_L=Q_R=Q=
\mbox{\rm diag.}[(2/3,-1/3,-1/3]$), in the presence of the strong
interactions ; the third line shows one of the lowest order terms in the
sector of the weak interactions. The typical physical processes to which
each term contributes are indicated under the braces. Each term is
modulated by a coupling constant:
$F_{0}^{2}$, $L_{10}$,... $C$...$g_{\underline{8}}$..., which encodes the
underlying dynamics responsible for the appearance of the
corresponding effective term. The evaluation of these couplings from the
underlying theory is the question we are interested in. The coupling 
$g_{\underline{8}}$ for
example, governs the strength of the dominant
$\Delta I=1/2$ transitions for
$K$ decays to leading order in the chiral expansion. 

There are two
crucial observations to be made concerning the relation of these low
energy constants to the underlying theory. The low--energy constants of
the Strong Lagrangian, like
$F_{0}^{2}$ and 
$L_{10}$,  are the
coefficients of the \underline{\it Taylor expansion}
of appropriate QCD Green's Functions. By contrast, the low--energy
constants  of the Electroweak Lagrangian, like e.g.
$C$ and $g_{8}$, are  
\underline{\it integrals} of appropriate QCD Green's Functions.
Their evaluation appears to be,
{\it a priori}, quite a formidable task because they require the
knowledge of Green's functions at all values of the euclidean
momenta; i.e. they require a precise {\it matching} of the {\it
short--distance} and the {\it long--distance} contributions to the
underlying Green's functions.
These two observations are generic in the case of the Standard
Model, independently of the $1/N_c$ expansion. The
large--$N_c$ approximation helps, however, because it restricts the
{\it analytic structure} of the Green's functions to be 
{\it meromorphic functions}; i.e., they
only have poles as singularities.

The $\QCD_{\infty}$ approach that we are proposing in order to compute a
specific coupling of the chiral electroweak Lagrangian consists of the
following steps:
\begin{enumerate}
\item{\it Identification of the relevant Green's functions.}
\item{\it Evaluation of the short--distance behaviour and the
long--distance behaviour of the relevant Green's functions.}
\item{\it Hadronic approximation of the underlying
Green's functions with a finite number of poles;} i.e., the minimum number
required to satisfy the  leading power fall--off at
short--distances, as well as the appropriate
$\chi$PT long--distance constraints.

We have checked this approach with the calculation of a few
low--energy observables:

{\em i) The electroweak $\Delta m_{\pi}$ mass
difference}~\cite{KPdeR98}. 

{\em ii) The hadronic vacuum polarization contribution to the
anomalous magnetic moment of the muon
$a_{\mu}$}~\cite{PdeR}.

{\em iii) The $\pi^{0}\ra e^{+}e^{-}$ and $\eta\ra \mu^{+}\mu^{-}$ decay
rates}~\cite{KPPdeR99}. 

These successful tests have encouraged us to pursue  a
systematic analysis of $K$--physics observables within the same
large--$N_c$ framework. So far, the following calculations have been made:

\begin{itemize}

\item{\it The $B_{K}$ Factor in the Chiral Limit}~\cite{PdeR00}.
\vspace*{-0.15cm}
\item{\it Weak Matrix Elements of the Electroweak Penguin Operators
$Q_7$ and $Q_8$}~\cite{KPdeR99,KPdeR01}.
\vspace*{-0.15cm}
\item{\it Hadronic Light--by--Light Scattering Contribution to the Muon
$g-2$}~\cite{KN02,KNPdeR02}.
\vspace*{-0.15cm}
\item{\it Electroweak Hadronic Contributions to the Muon
$g-2$}~\cite{KPPdeR02}.
\vspace*{-0.15cm} 
\item {\it Weak Matrix Elements of the Strong Penguin
Operators
$Q_4$ and $Q_6$}~\cite{HPdeR02}.

It appears now to be possible to apply the same techniques to the study of
rare
$K$ decays as well, and we shall illustrate this here, with a first
application to the process $K_L\ra\mu^{+}\mu^-$~\cite{GdeR03}.
 
\end{itemize}
\end{enumerate}

\section{RARE KAON DECAYS IN $\chi$PT}

In fact, it was at the Blois Conference in '98 where some of the first
applications of $\chi$PT to rare Kaon decays were reported.
The predicted invariant mass spectrum of the two gammas in the
mode $K_L\ra \pi^0 \gamma\gamma$, to lowest
order in the chiral expansion~\cite{EPdeRb87}~\cite{EdeR89} was discussed
prior to the earlier experimental measurements by the NA31 and E731
collaborations. Since then,
$\chi$PT has been confirmed as the appropriate framework to study rare
Kaon decays with many successful applications, ({\it see e.g.,
ref.~\cite{D'am01} for the latest conference review on the subject, where
further references can be found}.) 

Unfortunately, the predictive power of $\chi$PT is seriously limited at
present, because of the fact that several coupling constants of the higher
order terms in the electroweak chiral Lagrangian remain unknown. As a
result, many {\it predictions} at present rely on {\it models} which are
not clearly related to the underlying Standard Model theory. The analytic
approach which we have briefly summarized in the introduction offers,
however, an interesting possibility to make progress in this field.

Roughly speaking, from a theoretical point of view, there are three types
of $K$--decay modes:

\begin{itemize}
\item {The {\it golden modes} of $\chi$PT}

They correspond to processes which are fixed by $\chi$PT only. The
chiral loops are finite and there are no unknown counterterms at the
order one is working in the chiral expansion. Examples of that are
$K_S\ra\gamma\gamma$ and $K_L\ra\pi^0\gamma\gamma$ 

\item {The short--distance {\it golden modes}}

They directly test the structure of the Wilson coefficients, because the
relevant effective weak Hamiltonian for these processes is  of the unique
type: quark--current
$\times$ lepton current. The Wilson
coefficients have been calculated perturbatively with a sufficient degree
of accuracy which makes some of these modes an excellent laboratory of
possible Physics beyond the Standard Model. The paradigm mode of this type
is
$K_L\ra
\pi^0\nu\bar{\nu}$~({\it see e.g. ref.~\cite{Lit01} for a recent reviewÂ
on this mode}) 

\item {The  {\it mixed} modes}

This is, unfortunately, the largest class. Some modes are sensitive to
short--distance Physics, CP violation in  particular, like $K_L\ra\pi^0
e^+e^-$ and $\epsilon/\epsilon'$, but their calculation in the Standard
Model brings in unknown couplings of the chiral Lagrangian. It is here
where hard theoretical effort is mostly required.   
\end{itemize}


\section{$K_L\ra \mu^+ \mu^-$}

This is a decay, with both an interesting short--distance
component which is particularly sensitive to $V_{td}$, and a
long--distance contribution which has a large absorptive component from
the dominant
$\gamma\gamma$ intermediate state~\cite{MdeRS70}, illustrated in
Fig.~1. The present experimental rates are

\bea
\mbox{\rm Br}(K_{L}\ra \mu^{+}\mu^{-}) & = & (7.18\pm0.17)\times
10^{-9}\qquad {\mbox{\rm BNL--E871~\cite{BNLmu}}}\\
\mbox{\rm Br}(K_{L}\ra e^{+}e^{-}) & = & \left(8.7\begin{array}{c} +5.7\\
-4.1\end{array}\right)\times 10^{-12}\qquad {\mbox{\rm
BNL--E871~\cite{BNLel}}}
\eea

\vskip 3pc
\centerline{\epsfbox{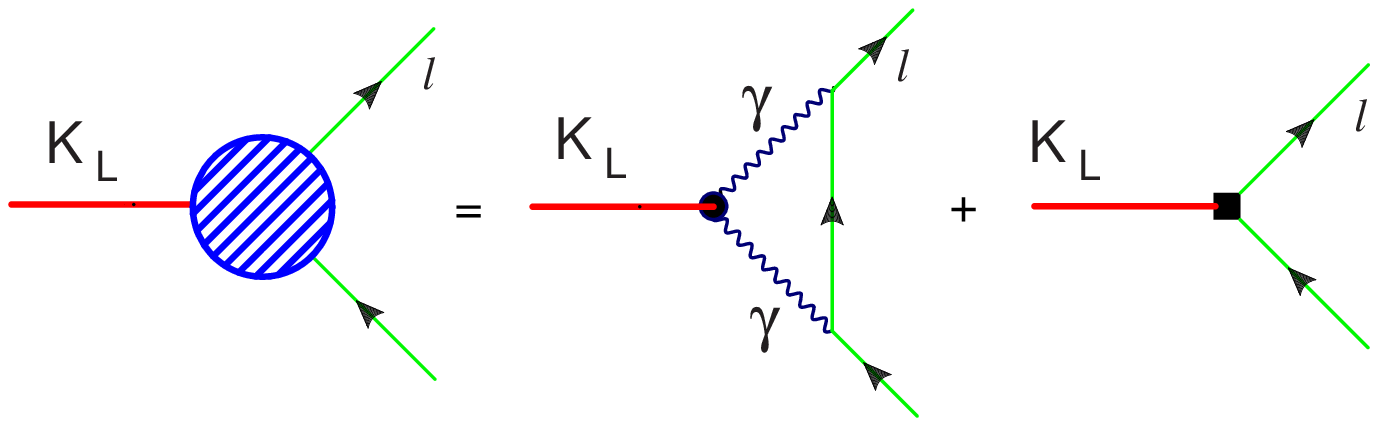}}
{{\bf Fig.~1} {\it Contribution to $K_L\ra\bar{l}l$ from the
$\gamma\gamma$ loop and local counterterms.}}
\vskip 3pc 

\noi
From a phenomenological point of view, it is convenient to normalize
the $K_L\ra \bar{l}l$ decay rate to the one of the $K_L\ra
\gamma\gamma$, which is also known experimentally [$\mbox{\rm
Br}(K_L\ra\gamma\gamma)=(5.86\pm0.17)\times 10^{-4}$]. Then
\be\lbl{brth}
\cR_{\bar{l}l}\equiv\frac{\Gamma\left(
K_L\ra\bar{l}l\right)}{\Gamma\left(
K_L\ra\gamma\gamma\right)}=2\beta\left(\frac{\alpha}{\pi}\right)^2
\left(\frac{m_{l}^2}{m_{K}^2}\right)\left\vert \cA_{\bar{l}l}
\right\vert^2\,,
\ee
with
\be
\Imm \cA_{\bar{l}l}=\frac{\pi}{2\beta_l}\log\left(
\frac{1-\beta_l}{1+\beta_l} \right)\,, \quad\annd\quad
\beta_l=\sqrt{1-\frac{4m_{l}^2}{M_K^2}}\,.
\ee
In fact, the contribution from the absorptive amplitude
$\Imm\cA_{\bar{\mu}\mu}$ almost saturates the observed experimental rate,
leaving very little room for the real part
\be
\vert\Ree\cA_{\bar{\mu}\mu}\vert^2<(7.1\pm0.2)\times 10^{-9}\,.
\ee

The real part of $\cA_{\bar{l}l}$ has a 
component from the
$\gamma\gamma$ loop which is divergent and requires a local
long--distance counterterm. The relevant couplings from the
long--distance effective Hamiltonian are the following
\be\lbl{LDH}  
\cH_{\eff}^{\mbox{\rm\tiny (LD)}}=\cC\left\{\frac{\alpha
N_c}{12\pi f_{\pi}}\epsilon_{\mu\nu\alpha\beta}
F^{\mu\nu}A^{\alpha}
(\partial^{\beta}K_L)+\left(\frac{\alpha}{\pi}\right)^2
\frac{\chi_{\mbox{\rm\tiny
LD}(\mu)}}{4f_{\pi}}(\partial_{\mu}K_L)\ 
\bar{l}\gamma^{\mu}\gamma_{5}l \right\}\,.
\ee
The first term provides the $K_L\gamma\gamma$--vertex interaction in
the loop diagram of Fig.~1, with an unknown coupling $\cC$ which can
however be fixed ({\it in modulus only}) from the observed
$K_L\ra\gamma\gamma$ decay rate:
\be\lbl{KLGG}
\Gamma(K_{L}\ra\gamma\gamma)=\frac{M_{K}^3}{64\pi}
\left(\frac{\alpha}{\pi}
\right)^2 \frac{1}{F_{0}^2}\mid
\cC\mid^2\,.
\ee

The second term is the long--distance contact term, which appears
at the same order in the chiral expansion as the contribution from the
$\gamma\gamma$ loop. The coupling
$\chi_{\mbox{\rm\tiny LD}}(\mu)$ is the quantity that one would like to
evaluate in the Standard Model.  In terms of this coupling, one finds

\be\lbl{ReALD}
{\Ree\cA}^{\mbox{\rm\tiny (LD)}}(\beta_l) = \chi_{\mbox{\rm\tiny
LD}}(\mu) +\frac{N_C}{3}\bigg[\,-\frac{5}{2}
+ 
\frac{3}{2}\ln\bigg(\frac{m_{\ell}^2}{\mu^2}\bigg)\,+\,C(\beta_l)\,\bigg]\,,
\lbl{amp}
\ee
where the function $C(\beta_l)$ corresponds to a finite 
three--point loop integral which is known, and
can be found in refs.~\cite{KPPdeR99,GDP98}. The divergence 
associated with this diagram has been 
renormalized within the ${\overline{\mbox{MS}}}$ minimal subtraction scheme 
of dimensional regularization. The logarithmic dependence on the 
renormalization scale $\mu$ 
displayed in the above expression is compensated by the scale dependence of 
the renormalized coupling $\chi_{\mbox{\rm\tiny
LD}}(\mu)$. Let us stress here that, in contrast 
with the usual situation in the purely mesonic sector, this scale dependence 
is not suppressed in the large--$N_C$ limit, since it does not arise from 
meson loops.

The amplitude $\cA_{\bar{l}l}$ also gets a tree level contribution from the
short--distance Hamiltonian~\cite{BB94}
\be\lbl{SDH}
\cH_{\eff}^{\mbox{\rm\tiny (SD)}}=\cdots
-\frac{\GF}{\sqrt{2}}\frac{\alpha}{\pi}\frac{2}{\ \sin^2\!\Theta_{W}}
\left[
\lambda_{c}Y_{NL}+
\lambda_{t}Y(x_t)\right] 
\left(\bar{s}\gamma_{\mu}\gL d\right)
\bar{l}\gamma^{\mu}\gL l+{\mbox{\rm h.c.}}\,,
\ee
where 
\be
\lambda_c=V_{cs}^{*}V_{cd}\,,\quad \lambda_{t}=V_{ts}^{*}V_{td}\,, \quad
x_{t}=\frac{m_{t}^2}{M_{W}^2}\,.
\ee
In the Standard Model,
this is the term in the effective Hamiltonian which  emerges after
integrating out the heavy degrees of freedom (Z, W; t, b, and c) in the
presence of the strong interactions. and which couples directly the quark
current
$\bar{s}\gamma_{\mu}\gL d$ to the lepton current  $\bar{l}\gamma^{\mu}\gL
l$. Here $Y_{NL}$ and $Y(x_t)$ are functions of the masses of the
integrated particles, which can be found in ref.~\cite{BB94}. 

The effect of the short--distance Hamiltonian \rf{SDH} in the amplitude
$\cA_{\bar{l}l}$ is to induce a shift in the $\chi_{\mbox{\rm\tiny
LD}}(\mu)$--coupling
\be
\chi_{\mbox{\rm\tiny
LD}}(\mu)\Ra \chi_{\mbox{\rm\tiny
LD}}(\mu)-\chi_{\mbox{\rm\tiny
SD}}\,,
\ee
with $\chi_{\mbox{\rm\tiny SD}}$ fixed by the relation
\be
-\frac{\GF}{\sqrt{2}}\frac{\alpha}{\pi}\frac{2}{\ \sin^2\!\Theta_{W}}
\Ree\left[
\lambda_{c} Y_{NL}+
\lambda_{t} Y(x_t)\right] =\frac{\alpha}{\pi}\ \cC\times
\chi_{\mbox{\rm\tiny SD}}\,.
\ee
Without loss of generality, we can fix $\mu$ at the $\rho$ mass and define
the parameter
\be\lbl{chic}
\chi_{\mbox{\rm\tiny eff}}=\chi_{\mbox{\rm\tiny
LD}}(M_{\rho})-\chi_{\mbox{\rm\tiny
SD}}
\ee
The branching ratio $\cR_{\bar{l}l}$ in Eq.~\rf{brth} is then a function of
only the unknown parameter $ \chi_{\mbox{\rm\tiny eff}}$, and the
predicted shapes for $l=\mu,e$ are as in Fig.~2. From the comparison with
the experimental value of $\cR_{\bar{\mu}\mu}$ (the horizontal band in
Fig.~3a) we conclude that
\be
3.9\le \chi_{\mbox{\rm\tiny eff}}\le 6.5\,.
\ee
The minimum of the predicted parabola in Fig.~2a corresponds to the lower
bound when
\be
\Ree\cA_{\bar{\mu}\mu}=0\quad\Ra\quad \chi_{\mbox{\rm\tiny eff}}=5.2.
\ee 

\vskip 1pc
\centerline{\epsfbox{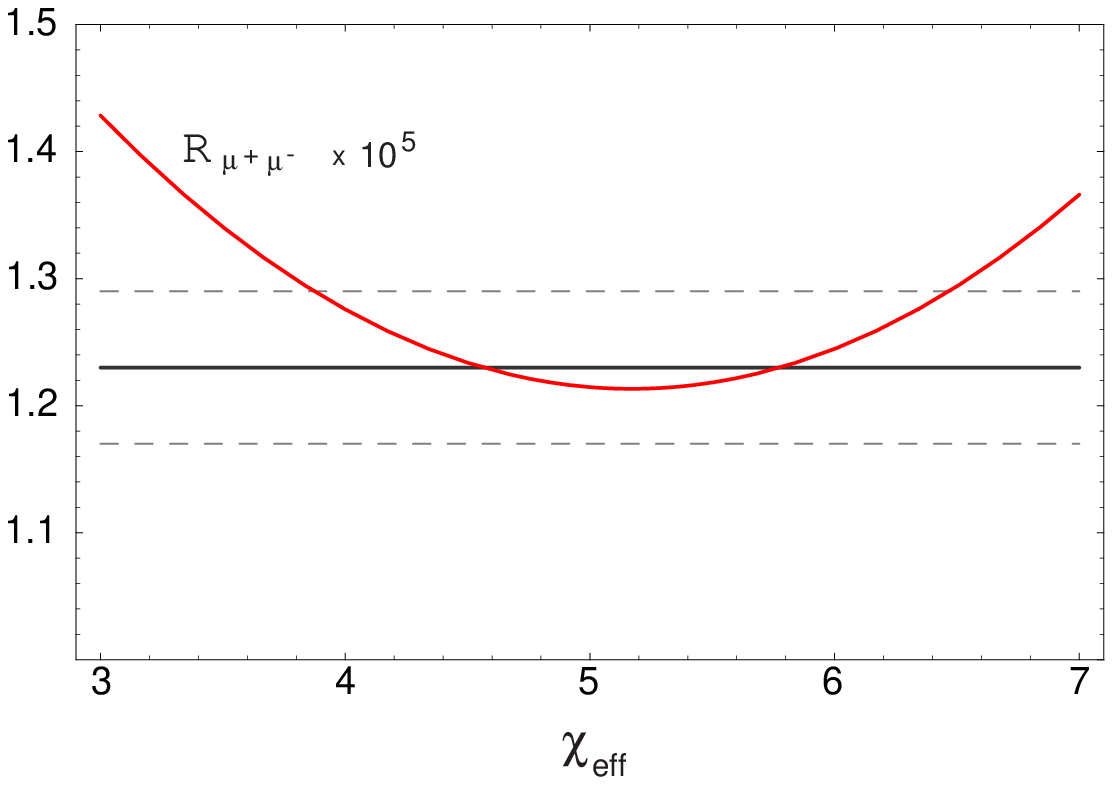}}
{{\bf Fig.~2a} {\it The predicted branching ratio $\cR_{\bar{\mu}\mu}$ as a
function of $\chi_{\mbox{\rm\tiny eff}}$. The horizontal lines correspond
to the present experimental result with errors.}}
\vskip 1pc 

\vskip 1pc
\centerline{\epsfbox{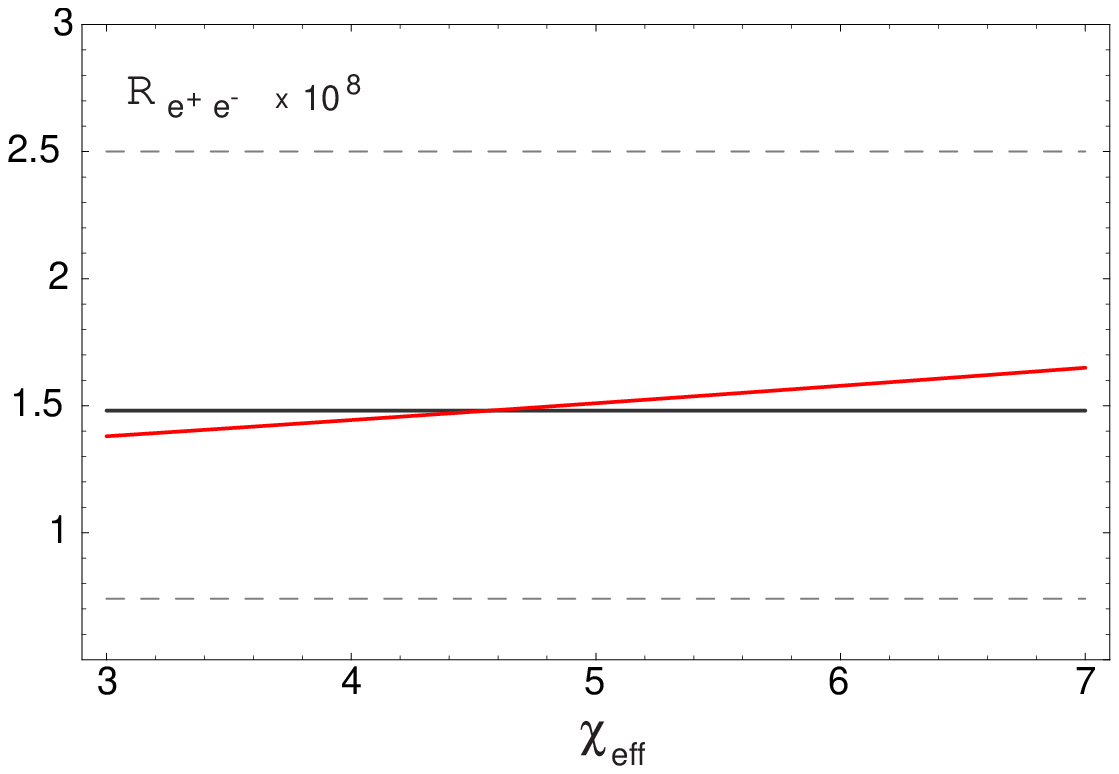}}
{{\bf Fig.~2b} {\it The predicted branching ratio $\cR_{\bar{e}e}$ as a
function of $\chi_{\mbox{\rm\tiny eff}}$. The horizontal lines correspond
to the present experimental result with errors.}}
\vskip 1pc 

\noi
Because of the small electron mass, the term $\log\frac{m_e^2}{M_{\rho}^2}$
dominates the r.h.s. of Eq.~\rf{ReALD}, and there is very little we can
learn on $\chi_{\mbox{\rm\tiny eff}}$ from the, otherwise, quite remarkable
experimental determination of $\cR_{\bar{e}e}$.

\subsection{EVALUATION OF $\chi_{\mbox{\rm\tiny
LD}}(\mu)$ IN THE LARGE--$N_c$ QCD FRAMEWORK }

The relevant Green's function for the process we are concerned with is the
four--point function

\bea\lbl{GF}
W_{\mu\nu}(q,p) & = & \lim_{l\ra 0}\int d^4x e^{iq\cdot x}\!\int d^4y
e^{ip\cdot y}\!\int d^4z e^{il\cdot z}\langle 0\vert
T\left\{J_{\mu}(x)J_{\nu}(y)\cP(0)\cL_{\mbox{\rm\tiny eff}}^{\Delta S=1}(z)
\right\}\!\!\vert 0\rangle  \\
 & = &
\frac{2}{3}\epsilon_{\mu\nu\alpha\beta}\ q^{\alpha}p^{\beta}
\frac{1}{(q+p)^2-M_{K}^2}\cW[q^2,p^2,(q+p)^2]\,, 
\eea
where, in the first line,
$J_{\mu}$ is the electromagnetic current;
$\cP=\bar{q}i\gamma_5 \frac{\lambda^{(6)}}{2}q$ the interpolating density
current operator of the
$K_{L}$--state and
$\cL_{\mbox{\rm\tiny eff}}^{\Delta S=1}$ the strangeness changing chiral
effective Lagrangian.
The function $\cW[q^2,p^2,(q+p)^2]$ governs both the $K_L\ra
\mu^{+}\mu^{-}$  and the $K_L\ra\gamma\gamma$ decay rates. In particular,
the relation to the constant
$\cC$ introduced in Eqs.~\rf{LDH} and \rf{KLGG} is as follows 
\be
\cW[0,0,M_{K}^2]=\frac{N_c}{8\pi^2}\ \frac{\vert\stern\vert}{F_0^2}\ \cC\,,
\ee
while the evaluation of the $K_L\ra
\mu^{+}\mu^{-}$ amplitude requires the knowledge of the
function $\cW[q^2,q^2,M_K^2]$ at all values $0\le
-q^2\le\infty$.

  It is well known that the lowest order evaluation of the constant
$\cC$ in $SU(3)_L \times SU(3)_R$ $\chi$PT gives a contribution
which is proportional to the lowest order Gell-Mann Okubo mass relation: 
$3M_{\eta}^2-4M_{K}^2+M_{\pi}^2=0$. The first non--trivial contribution to
$\cC$ comes from  chiral loops and a combination of undetermined couplings
of the effective
$\cL_{\mbox{\rm\tiny eff}}^{\Delta S=1}$ Lagrangian at $\cO(p^4)$. This is
why it has become so difficult to evaluate 
$K_{L}\ra\gamma\gamma$, and hence $K_L\ra\mu^+\mu^-$ in the Standard Model.
We do not attempt to do a calculation of $\cC$, but we shall use a crucial
observation which, to leading non--trivial order in the
chiral expansion, relates the residue of the short--distance behaviour
($-q^2\ra\infty$) of the function $\cW[q^2,q^2,M_K^2]$ to the constant
$\cC$, known in modulus from the observed $K_{L}\ra\gamma\gamma$ decay
rate~\footnote{It is in this sense that we differ from the large--$N_c$
inspired $U(3)_L\times U(3)_R$ phenomenological analysis of
ref.~\cite{GDP98}. (See also the discussion in \cite{KPdeR00}.)}.  The
relation is the following~\footnote{Notice the the short--distance
behaviour obtained here is in disagreement with the {\it hadronic ansatz}
proposed in ref.~\cite{AIP98}.} 
\be\lbl{mainres}
\lim_{-q^2\ra\infty}\cW[q^2,q^2,M_{K}^2]=\frac{\vert\stern\vert}{-q^2}\ 
\left(\cC+
\frac{8}{9}\ 4L_{5}\
\frac{M_K^2}{F_{0}^2}\ \frac{M_K^2-M_{\pi}^2}{M_{K}^2-M_{\eta}^2}
\right)\,,
\ee
and is based on the following facts:

\begin{itemize}
\item The first non--trivial contribution to the constant $\cC$ in $\chi$PT
and to leading order in the $1/N_c$ expansion, comes from the bosonization
of the
$T$--product in Eq.~\rf{GF}, with the contribution to $J_{\mu}(x)J_{\nu}(y)$
induced by the local \underline{one--Goldstone} contribution from the
$\cO(p^4)$ Wess--Zumino Lagrangian, the $\cO(p^2)$ bosonization of $\cP(0)$
and the problematic
$\cO(p^4)$ contribution from $\cL_{\mbox{\rm\tiny eff}}^{\Delta
S=1}(z)$~\footnote{We have identified the combination of $\cO(p^4)$
couplings which contribute~\cite{GdeR03}.}

\item
When performing the operator product expansion of the two electromagnetic
currents, the leading contribution is governed by the local operator:
$\bar{q}Q^2 \gamma^{\beta} q$, which also has to be bosonized in
terms of \underline{one--Goldstone} states and  produces the
\underline{same} combination of fields as the one emerging from the
Wess--Zumino
$J_{\mu}(x)J_{\nu}(y)$ product discussed in the previous item. The
resulting
$T$--product of this $\bar{q}Q^2 \gamma^{\beta} q$--operator, with 
 the remaining $\cP(0)$ and
$\cL_{\mbox{\rm\tiny eff}}^{\Delta S=1}(z)$ operators, has to be evaluated
at long--distances.  To first non--trivial order, there are  two pieces
which contribute to its bosonization: the one from $\bar{q}Q^2
\gamma^{\beta} q$ at $\cO(p^2)$ with  $\cP(0)$ at $\cO(p^2)$ and
$\cL_{\mbox{\rm\tiny eff}}^{\Delta S=1}(z)$ at $\cO(p^4)$  produces a term
proportional to the \underline{same} contribution discussed in the previous
item; the other one from 
$\bar{q}Q^2 \gamma^{\beta} q$ at $\cO(p^4)$, which is proportional
to the
$L_5$ coupling, times  $\cP(0)$ at $\cO(p^2)$ and 
$\cL_{\mbox{\rm\tiny eff}}^{\Delta
S=1}(z)$ also at $\cO(p^2)$. Fortunately, this second contribution brings in
known couplings of the chiral Lagrangian. 

\item The other possible contributions to $\cW[0,0,M_K^2]$ and
$\lim_{-q^2\ra\infty}\cW[q^2,q^2,M_K^2]$, at the same order in the chiral
expansion, either give vanishing contributions, or are proportional to the
Gell-Mann Okubo combination of masses.  However, higher order contributions
in the chiral expansion may very likely destroy this dynamical symmetry
observed at the first non--trivial order. This is why our main result in
Eq.~\rf{mainres} is only valid to first non--trivial order in the chiral
expansion. Furthermore, the effect of chiral loops, subleading in the
$1/N_c$ expansion are also neglected.
\end{itemize}

The procedure to evaluate the effective coupling $\chi_{\mbox{\rm\tiny
LD}}(\mu)$ is now entirely similar to the one discussed in
ref.~\cite{KPPdeR99} with the result
\be
\chi_{\mbox{\rm{\tiny LD}}}(M_V)  =  \frac{11}{12}N_c-4\pi^2
\frac{F_{0}^2}{M_{V}^2}\left(1\pm 
\frac{8}{3}\frac{M_K^2-m_{\pi}^2}{F_{0}^2}\
\frac{4L_{5}}{1.19\pm 0.16} \right)\,,
\ee
where the first $\pm$ sign comes from the fact that the constant $\cC$ is
only known in modulus. This implies a two--fold result for the effective
coupling $\chi_{\mbox{\rm{\tiny eff}}}$ in Eq.~\rf{chic} which, using the
short--distance evaluation $\chi_{\mbox{\rm{\tiny
SD}}}=\pm(1.8\pm0.1)$, results in 
\be
\chi_{\eff}=\left.\begin{array}{l} 2.33\pm 0.9 -(1.8\pm 0.1)= 
0.5\pm 0.9
\\  2.03\pm 0.9 + (1.8\pm 0.1)= 3.8\pm 0.9 \end{array}\right\}
\ee

\noi
The corresponding branching ratios for the two solutions are shown in
Figure~3 below

\vskip 1pc
\centerline{\epsfbox{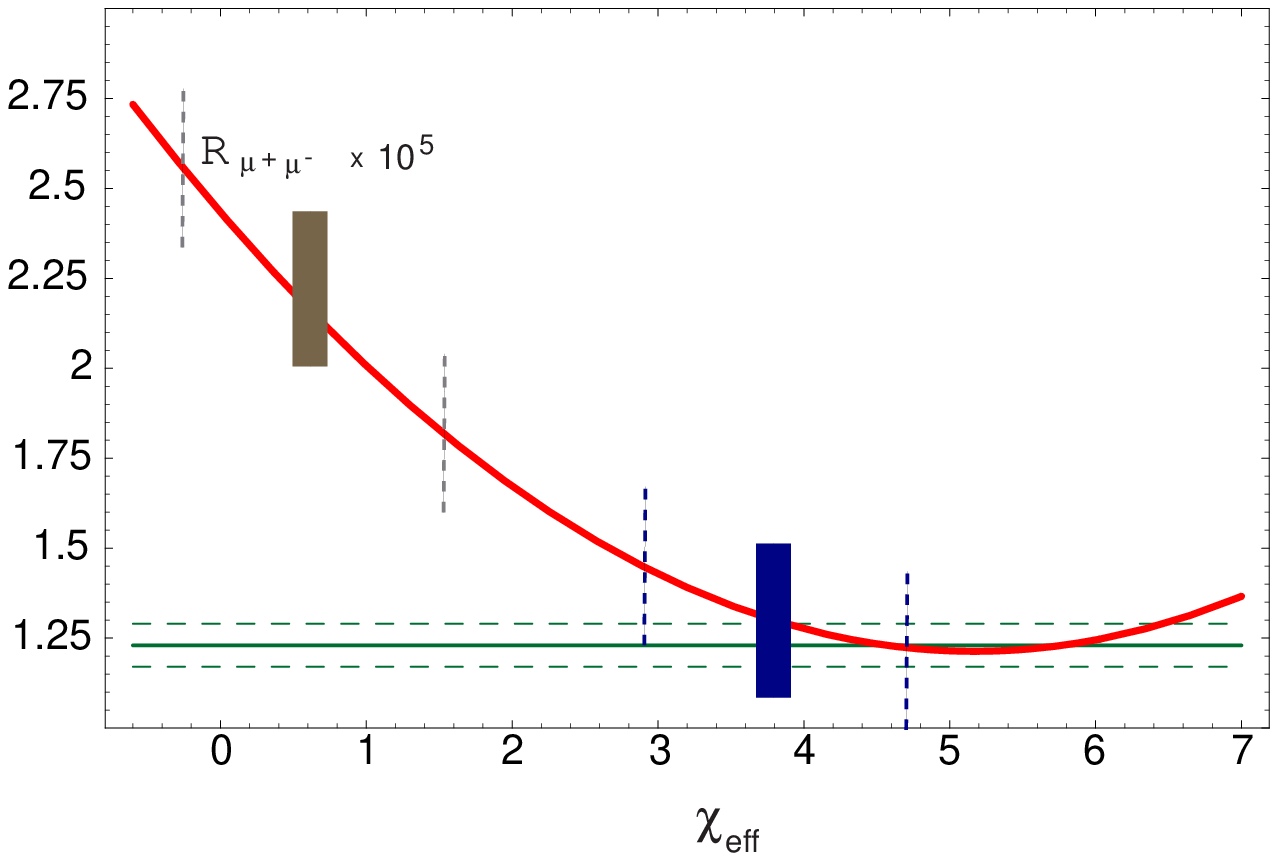}}
{{\bf Fig.~3} {\it Branching ratio versus $\chi_{\eff}$. The horizontal
band corresponds to the experimental value with errors . The predicted
solutions are the two solid vertical bands with vertical error bars.}}
\vskip 1pc 

The higher solution is in perfect agreement with the experimental
determination, the lower one is in slight difficulty. 

It is very likely that  considerations similar to the ones illustrated here
in the case of $K_L\ra\mu^{+}\mu^-$, may also  apply to other rare
$K$--decay processes and open, therefore, a new theoretical perspective 
in the field of rare $K$ decays.

\section*{Acknowledgments}
We wish to thank Samuel Friot, Marc Knecht and Santi Peris for helpful
discussions. This work has been partially supported by the TMR, EC-Contract
No. HPRN-CT-2002-00311(EURIDICE).


\end{document}